\documentclass[aps,pra,twocolumn,superscriptaddress,floatfix,times,showpacs]{revtex4-1}
\usepackage{graphicx}
\usepackage{graphics}
\usepackage{bm,dsfont}
\usepackage{amsmath, amsthm, amssymb,color}
\usepackage{epstopdf}
\usepackage{dcolumn,txfonts}

\theoremstyle{definition}

\definecolor{Blue}{rgb}{0,0,1}
\definecolor{Red}{rgb}{1,0,0}
\definecolor{Green}{rgb}{0,1,0}
\definecolor{Purp}{rgb}{.2,0,.2}
\definecolor{white}{rgb}{1,1,1}

\begin{document}

\newcommand{\field}[4] {{\mathbf #1}{#2} ({\bf #3} , #4) }
\newcommand{\Fq}[2]{ {\mathbf #1}{#2}({\bf q},t )}
\newcommand{\Fr}[2]{ {\mathbf #1}{#2}({\bf r},t)}
\newcommand{\Balpha}{\boldsymbol\alpha}
\newcommand{\Bbeta}{\boldsymbol\beta}
\newcommand{\ddt}{\frac{\partial}{\partial t}}
\newcommand{\ket}[1]{\left| #1 \right\rangle}
\newcommand{\bra}[1]{\left \langle #1 \right |}
\newcommand{\braket}[1]{\left \langle #1 \right \rangle}

\newcommand{\A}{\mathcal{A}}
\newcommand{\s}{\mathcal{S}}
\newcommand{\e}{\mathcal{E}}
\newcommand{\se}{\mathcal{SE}}
\newcommand{\rs}{\rho^{\mathcal{S}}}
\newcommand{\re}{\rho^{\mathcal{E}}}
\newcommand{\rse}{\rho^\se}
\newcommand{\tr}{\mbox{Tr}}
\newcommand{\tre}{{\rm Tr}_\mathcal{E}}
\newcommand{\trs}{{\rm Tr}_\mathcal{S}}

\title{Vanishing quantum discord is \emph{not} necessary for completely-positive maps}

\author{Aharon Brodutch}
\email{abrodutc@uwaterloo.ca}
\affiliation{Institute for Quantum Computing and Department of Physics and Astronomy, University of Waterloo, Waterloo ON N2L 3G1, Canada}

\author{Animesh Datta}
\email{a.datta1@physics.ox.ac.uk}
\affiliation{Clarendon Laboratory, University of Oxford, Parks Road, Oxford, UK, OX1 3PU}

\author{Kavan Modi}
\email{kavan@quantumlah.org}
\affiliation{Clarendon Laboratory, University of Oxford, Parks Road, Oxford, UK, OX1 3PU}
\affiliation{Centre for Quantum Technologies, National University of Singapore, Singapore}

\author{\'{A}ngel Rivas}
\email{angelitorv@gmail.com}
\affiliation{Departamento de F\'isica Te\'orica I, Facultad de Ciencias F\'isicas, Universidad Complutense, 28040 Madrid, Spain}

\author{C\'esar A. Rodr\'iguez-Rosario}
\email{cesar.rodriguez@bccms.uni-bremen.de}
\affiliation{Dept. of Chemistry and Chemical Biology, Harvard University, Cambridge Massachusetts 02138, USA}
\affiliation{Bremen Center for Computational Materials Science, University of Bremen, Bremen, Germany}

\date{\today}
\begin{abstract}
The description of the dynamics of a system that may be correlated with its environment is only meaningful within the context of a specific framework. Different frameworks rely upon different assumptions about the initial system-environment state. We reexamine the connections between complete-positivity and quantum discord within two different sets of assumptions about the relevant family of initial states. We present an example of a system-environment state with non-vanishing quantum discord that leads to a completely-positive map. This invalidates an earlier claim on the necessity of vanishing quantum discord for completely-positive maps. In our final remarks we discuss the physical validity of each approach.
\end{abstract}
\pacs{03.67.-a, 03.65.Ud, 03.65.Yz}

\maketitle

\section{Introduction}
The open quantum system formalism is the standard tool used to model the decoherence and thermalization of quantum systems~\cite{Angelbook}. The total state of the system ($\s$) and its environment ($\e$) is initially described by the density matrix $\rho^{\se}(0)$, which evolves unitarily to $\rse(t)$. The dynamics of the density matrix of $\s$ is obtained by averaging over the degrees of freedom of $\e$. It is customary to assume that the initial system-environment ($\se$) state is uncorrelated. This assumption simplifies the mathematical structure of the dynamical map that describes the evolution of a family of system states and leads to a clear method for quantum process tomography. However, researchers have been trying to understand how to construct a dynamical map when there are initial system-environment correlations~\cite{NCP, SL, Pechukas, buzek, JSS, CTZ, Laine:2011hc, MSS, Hutter:2011tc, Devi:2011wt}.

When initial $\se$ correlations are present, even the relevant questions regarding the reduced dynamics require careful consideration. A general question such as `\emph{Can we define a completely-positive map from a family of initial states $\{\rho^\s(0)\}$ to final states $\{\rho^\s(t)\}$}?' is not well posed. For instance, if one regards a \emph{single} initial state then there always exists a completely-positive (CP) transformation that connects the initial state to the final state. If on the other hand, one does not specify anything about the correlations then the answer is generally `no' since the reduced dynamics can be non-linear~\footnote{For any initial system state, consider for example an environment containing many copies of that initial system state. See also~\cite{Ternononlinear}.}. The question only becomes meaningful when one considers a specific framework for defining the relevant initial system-environment states.

Recently, a relation between completely-positive maps and quantum discord has been put forth~\cite{NCP, SL}. In~\cite{NCP}, it is shown that if the initial $\se$ state has vanishing quantum discord, then the dynamics of $\s$ can be described by a CP map. In~\cite{SL} Shabani and Lidar made a strong claim: \emph{The reduced dynamics of a system is completely-positive, for any coupling with the environment, only if the initial system-environment state has vanishing quantum discord as measured by the system}. 

Thus, the relation to quantum discord seemed to settle the question about CP character of reduced dynamical maps. This connection was also used to justify much of the work on quantum discord since it presumably plays a significant role in open systems dynamics. At the time of publication, there were only a handful of results connecting discord to interesting physical scenarios and this connection to CP maps was seen as a significant motivating factor for studying discord. We should note, however  that despite the  vast interest in this connection it was never used for any substantial results.

Here we show that underlying assumptions in~\cite{SL} limit the generality of the constructed reduced dynamics. To do this we consider two different paradigms to ask the question \emph{Given any unitary evolution for the system-environment, can we define a completely-positive map taking a family of initial states $\{\rho^\s(0)\}$ to final states $\{\rho^\s(t)\}$?}

\begin{enumerate}
\item[(i.)] A framework built with a family of initial $\se$ states $\{\rho^\se(0) = \sum_{i,j} \rho_{ij} \sigma_{ij}^\s \otimes \phi_{ij}^\e \}$ spanned by variables $\{\rho_{ij}\}$, while operators $\{\sigma_{ij}^\s \otimes \phi_{ij}^\e\}$ are kept fixed.

\item[(ii.)] A framework built with a family of initial $\s$ states belonging to the consistency domain of a linear map $\A$, called an assignment map, that maps operators on $\s$ to operators on $\se$. A state is said to belong to the consistency domain when it satisfies $\{ \rho^\s(0) = \tr_\e \left(\A[ \rho^\s(0)]\right)\}$.
\end{enumerate}

In (i) the dynamical map will depend directly on the family of initial $\se$ states. While, in (ii) it will depend on the assignment map. Additionally, in both frameworks the dynamical map is also a function of the unitary evolution.

 It is the aim of this article to re-examine the result of~\cite{SL} using two frameworks stated above. We show that in both cases CP maps can describe the dynamics of $\s$ for any unitary even in a case where the initial $\se$ state is discordant. We go on to discuss the limitations of the first framework and demonstrate that it suffers from a number of fundamental issues. Finally, in concluding remarks we discuss the relation to quantum process tomography.

We begin with framework (i) in Sec.~\ref{SectionSL} with a brief review a crucial part of the argument given in~\cite{SL}. In Sec.~\ref{SectionCE} we construct an example of an initial $\se$ state that has non-vanishing discord~\cite{Ollivier01a}, yet the reduced dynamics of $\s$ is given by a CP map independent of the coupling to $\e$. In Sec. IV, we discuss the standard argument that a map that depends on the state cannot be considered linear. We then argue how such nonlinearity can be (mathematically) avoided when one uses assignment maps. We conclude in Sec.~\ref{conclusions} by highlighting how unspecified assumptions can affect the formulation of questions regarding complete-positivity and initial correlations. We finally discuss what is operationally meaningful when dealing with open dynamics with initial $\se$ correlations.

\section{Background}\label{SectionSL}

The main result of this article is in the context of the main result of~\cite{SL}, therefore we begin by sketching that result. The argument in~\cite{SL} begins by writing the $\se$ density matrix as
\begin{gather}
\label{slclass}
\rse(0) = \sum_{ij} \rs_{ij} \ket{i}\bra{j} \otimes \phi^\e_{ij},
\end{gather}
where $\{\phi^\e_{ij}\}$ are fixed operators on the spaces $\e$ such that if $\tr[\phi^\e_{ij}] \neq 0$ then $\tr[\phi^\e_{ij}] = 1$, and $\braket{i|j}=\delta_{ij}$. The proof of the main statement of~\cite{SL} relies on its Lemma 1 and Lemma 4. There the authors considered the singular-value decomposition of the operators on $\e$
\begin{gather}
\phi^\e_{ij} = \sum_\alpha \lambda_{ij}^\alpha \ket{x_{ij}^\alpha} \bra{y_{ij}^\alpha},
\end{gather}
here $\lambda_{ij}^\alpha$ are the singular values, so that they are real and positive. The reduced dynamics of $\s$ can be expressed as
\begin{align}
\rs(t) \equiv& \Phi[\rs(0)] = \tr_\e[U \rse(0) U^\dag]\nonumber \\
=&\sum_{kij\alpha} \sqrt{\lambda^\alpha_{ij}}
\bra{\varphi_k}U\ket{x_{ij}^\alpha} \rs_{ij} \ket{i} \bra{j}
\sqrt{\lambda^\alpha_{ij}} \bra{y_{ij}^\alpha} U^\dag\ket{\varphi_k}\nonumber \\
=&\sum_{kij\alpha} V_{kij}^\alpha \left(\rs_{ij} \ket{i} \bra{j} \right) W_{kij}^{\alpha\dag},
\end{align}
where $V_{kij}^\alpha =\sqrt{\lambda^\alpha_{ij}} \braket{\varphi_k \left| U \right| x^\alpha_{ij}}$ and $W_{kij}^\alpha = \sqrt{\lambda^\alpha_{ij}} \braket{\varphi_k \left| U \right| y^\alpha_{ij}}$. Here $U$ denotes some joint-unitary dynamics and $\{\ket{\varphi_k}\}$ is a basis of $\e$. Then they made use of operators $P_m=\ket{m}\bra{m}$ such that $\sum_{mn} P_m \rs(0) P_n = \rs(0)$.
\begin{align}
\rs(t) =& \sum_{kij\alpha} V_{kij}^\alpha P_i \left(\sum_{mn} \rs_{mn} \ket{m} \bra{n} \right) P_j W_{kij}^{\alpha\dag} \nonumber \\
 =& \sum_{kij\alpha} E_{kij}^\alpha \rs(0) F_{kij}^{\alpha\dag}.\label{mapproj}
\end{align}
where $E_{kij}^\alpha = V_{kij}^\alpha P_i$ and $F_{kij}^\alpha = W_{kij}^\alpha P_j$. The explicit form of the map, $\Phi$, is therefore (as shown in Eq.~(14) of~\cite{SL}):
\begin{gather}\label{mapintheslway}
\Phi[\rs(0)]=\sum_{kij\alpha} E_{kij}^\alpha \rs(0) F_{kij}^{\alpha\dag}.
\end{gather}
The map $\Phi$ is defined on the family of system states $\{\rs(0)\}$ spanned by parameters $\{\rs_{ij}\}$ in Eq.~\eqref{slclass}.

\emph{Quantum discord.} A state with vanishing discord as measured by $X$,~\cite{Ollivier01a} can be written as $\rho^{XY}=\sum_i p_i \ket{\chi_i}\bra{\chi_i}^X \otimes \rho_i^Y$, where $\{p_i\}$ are probabilities, $\{\ket{\chi_i}\}$ form an orthonormal rank-one basis on the space $X$ and $\{\rho_i^Y\}$ are density matrices. Note that $\{\ket{\chi_i}\}$ is the eigenbasis of state $\rho^{X}$.

Suppose the initial $\se$ state has vanishing discord as measured by $\s$. Thus, we may write Eq.~\eqref{slclass} as
\begin{gather}\label{slclasszd}
\rse(0) = \sum_{ii} \rs_{ii} \ket{\chi_i}\bra{\chi_i} \otimes \phi^\e_{ii},
\end{gather}
where the basis $\{\ket{\chi_i}\}$ is the eigenbasis of $\s$. Then the dynamical map $\Phi$ could be constructed by choosing $\phi^\e_{ij}=0$, $\phi^\e_{ii}=\rho^\e_{ii}$, $P_i = \ket{\chi_i} \bra{\chi_i}$, so that we have $\ket{x_{ij}^\alpha} = \ket{y_{ij}^\alpha}$, operators $E^\alpha_{kij}=F^\alpha_{kij}$ and the map, with Eq.~\eqref{mapintheslway}, becomes the Kraus representation of a CP map~\footnote{Note that for the special case where $\phi_{ij} = \phi_{ji} = \phi_{ii} = \phi_{jj}$ for some $i,j$ there is more freedom in choosing the basis of $\s$ within that subspace.}. This map is defined on the family of system states spanned by parameters $\{\rs_{ii}\}$ in Eq.~\eqref{slclasszd}. The main claim in~\cite{SL} is that the converse is also true, i.e., a family that includes a discordant state cannot generate a reduced CP dynamics.

\section{Vanishing discord is not necessary for CP maps}\label{SectionCE}

We now construct a counter example to the central claim of~\cite{SL}, that vanishing discord is not necessary for a reduced CP map. Consider the $\se$ state
\begin{align}\label{ex1}
\rho^{\se}_{\rm c}(0)=&\frac{p}{3}\left(\ket{0}\bra{0} \otimes \rho_0^{\e} +\ket{1}\bra{1} \otimes \rho_1^{\e} +\ket{+}\bra{+} \otimes \rho_+^{\e}\right)
\nonumber \\ &+ \sum_{i=2}^n p_i\ket{i}\bra{i} \otimes \rho_i^{\e}.
\end{align}
with the probability $p=1-\sum_{i=2}^{n}p_i$. This state has non-vanishing discord, except for the special cases of $p=0$ or $\rho^{\e}_1=\rho^{\e}_0=\rho^{\e}_+$, otherwise $[\rho^{\s}_{\rm c}(0) \otimes \mathds{1}^{\e},\rho^{\se}_{\rm c}(0)]\ne0$ which is a simple way to test for states with non-vanishing discord~\cite{faca}. We rewrite this state $\rho^{\se}_{\rm c}(0)$ as:
\begin{gather}\label{slclass2}
\rho^{\se}_{\rm c}(0)=\sum_{i} \rs_{ii} \ket{\psi_i}\bra{\psi_i} \otimes \phi^\e_{ii},
\end{gather}
where, remarkably, this decomposition is not like Eq.~\eqref{slclasszd} because we do not require an orthonormal basis of $\s$; i.e., $\braket{\psi_i|\psi_j} \neq \delta_{ij}$. Then we construct a map for this state for some joint dynamics $U$. By simple relabeling,
\begin{align}
\ket{\psi_{+}}=\ket{+}, \; \mbox{and} \; \ket{\psi_{j}} = \ket{j} \ \mbox{for $j=0,1,2,\ldots$}
\end{align}
that correspond to
\begin{align}\label{basisforcpmap}
\phi^\e_{++} = \rho^{\e}_+, \; \mbox{and} \; \phi^\e_{jj} = \rho^{\e}_{j} \ \mbox{for $j=0,1,2,\ldots$}
\end{align}
Note that each of the operators above is Hermitian and has unit trace. Taking the spectral decomposition of each we get
\begin{align}
\phi^\e_{++}&=\sum_\alpha \lambda^\alpha_+ \ket{x^\alpha_+}\bra{x^\alpha_+}, \label{OPE1}\\
\phi^\e_{jj}&=\sum_\alpha \lambda^\alpha_j \ket{x^\alpha_j}\bra{x^\alpha_j}\ \label{OPE2} \mbox{for $j=0,1,2,\ldots$}
\end{align}
Thus the state of $\s$ at a later time under the joint-dynamics $U$ will be
\begin{align}\label{CPCE1}
\rho^{\s}_{\rm c}(t) =&\tre \left[ U \rho^{\se}_{\rm c}(0) U^\dag \right] \\
=& \sum_{k,\alpha} \big\{\frac{p}{3} V_{k0}^\alpha \ket{0}\bra{0} V_{k0}^{\alpha\dag} + \frac{p}{3}V_{k1}^\alpha \ket{1}\bra{1} V_{k1}^{\alpha\dag} \nonumber \\
&+ \frac{p}{3}V_{k+}^\alpha \ket{+}\bra{+} V_{k+}^{\alpha\dag} + \sum_{j=2}^n p_j V_{k j}^\alpha \ket{j}\bra{j} V_{kj}^{\alpha\dag} \big\}, \nonumber
\end{align}
where $V_{k+}^\alpha = \sqrt{\lambda^\alpha_+} \braket{\varphi_k \left| U \right| x_+^\alpha}$ and $V_{kj}^\alpha= \sqrt{\lambda^\alpha_j}\braket {\varphi_k \left| U \right| x_j^\alpha}$ for $j=0,1,2,\dots$, and $\{\ket{\varphi_k}\}$ forms a complete basis of $\e$. We can convert this into a form that is clearly CP
\begin{align}
\rho^{\s}_{\rm c}(t)=& \sum_k \big\{ E_{k+0}^\alpha \rho^{\s}_{\rm c}(0) E_{k+0}^{\alpha\dag} + E_{k+1}^\alpha \rho^{\s}_{\rm c}(0) E_{k+1}^{\alpha\dag} \nonumber \\
&+ \sum_{j=0}^n E_{k j}^\alpha \ \rho^{\s}_{\rm c}(0) \ E_{k j}^{\alpha\dag} \big\}, \label{CPCE}
\end{align}
here we have
\begin{align}
&E_{k+0}^\alpha=V_{k+}^\alpha M_{+0}, \; E_{k+1}^\alpha=V_{k+}^\alpha M_{+1},\nonumber \\
&E_{k j}^\alpha=V_{k j}^\alpha M_{j} \quad \mbox{for $j=0,1,2,\ldots$}
\end{align}
The matrices $\{M\}$ are given by
\begin{align}
&M_{0} = \sqrt{\tfrac{2}{3}} \ket{0}\bra{0}, \; M_{1} = \sqrt{\tfrac{2}{3}} \ket{1}\bra{1}, \label{POVM1}\\
&M_{+0} = \sqrt{\tfrac{1}{3}} \ket{+}\bra{0}, \; M_{+1} = \sqrt{\tfrac{1}{3}} \ket{+}\bra{1}, \label{POVM2}\\
&M_{j} = \ket{j}\bra{j} \quad \text{for } j \ge 2. \label{POVM3}
\end{align}
For clarity, note that
\begin{align}
& M_{j} \rho^\s_{\rm c}(0) M_{j}^\dag = \tfrac{2 p}{3} \ket{j}\bra{j} \;
{\rm for} \ j=0,1, \\
& M_{+0} \rho^\s_{\rm c}(0) M_{+0}^\dag = M_{+1} \rho^\s_{\rm c}(0) M_{+1}^\dag =\tfrac{p}{6}\ket{+}\bra{+}, \\
& M_{j} \rho^\s_{\rm c}(0) M_{j}^\dag = p_j \ket{j}\bra{j} \; {\rm for} \ j = 2,\dots,n.
\end{align}

Finally, we can write Eq.~\eqref{CPCE} as $\rho^\s_{\rm c}(t)=\sum_{\varepsilon} K_{\varepsilon} \; \rho^\s_{\rm c}(0) \; K_{\varepsilon}^\dag$, where $\varepsilon$ groups together all the indices $\varepsilon:=\{k,j,\alpha,+1,+0\}$, therefore the dynamical map in Eq.~\eqref{CPCE} is CP. It is also easy to check that the completeness relation $\sum_{\varepsilon} K_{\varepsilon}^\dag K_{\varepsilon}=\mathds{1}$ is satisfied. The CP map constructed here is defined on the family of system states spanned by parameters $\{p, \; p_i\}$; this family includes discordant states. We conclude that it is possible to describe the reduced evolution of the class of states in Eq.~\eqref{ex1} with a CP map for any $\se$ coupling ($U$). This contradicts the claim of~\cite{SL} that of the necessity of vanishing discord for CP dynamics.

\subsection{Discussion of the counter example}\label{discussion}

The construction of the map in~\cite{SL} relies in the decomposition of the $\se$ state as Eq.~\eqref{slclass}, with orthonormal elements of $\s$: $\braket{i|j} = \delta_{ij}$. A state written in the form of Eq.~\eqref{slclass} has a diagonal representation when it has vanishing discord. However, we have written the initial state of our counter example in a diagonal form in Eq.~\eqref{slclass2}. We have constructed a map relaxing the orthogonality of the basis $\{\ket{\psi_i}\}$ in Eq.~\eqref{slclass2}. If we use the orthogonal decomposition Eq.~\eqref{slclass} for the state in Eq.~\eqref{ex1} we do not obtain a CP map, which is in accordance with the result in~\cite{SL}. Nevertheless, there is no good reason to prefer either decomposition of Eqs.~\eqref{slclass} or \eqref{slclass2}, since both represent the same $\se$ state.

Furthermore, the construction in~\cite{SL} relies on operators $\{P_m\}$, in Eq.~\eqref{mapproj}, to be rank-one, Hermitian projections on $\s$. The matrices ${M}$ of Eqs.~\eqref{POVM1}, \eqref{POVM2} and \eqref{POVM3} are not Hermitian, rank-one projections as in Eq.~\eqref{mapproj}. Incidentally, these matrices are elements of a positive operator valued measure (POVM) $\left\{ \Pi_\nu := M_\nu^\dag M_\nu \right\}$ (with $\nu \in\{+(0),+(1),0,1,2,\ldots\}$), that leave $\rs_{\rm c}(0)$ invariant.

In addition, we stress that in~\cite{SL} the specific construction of the maps was used to establish the condition $\Phi[\ket{i} \bra{j}] =\tre[U \ket{i} \bra{j} \otimes \phi^\e_{ij} U^\dagger]$. Notably, this condition only holds to compute Choi matrices from a maximally entangled state written in the basis $\{\ket{i}\}$ (i.e., the same basis as the decomposition of Eq.~\eqref{slclass}). This already suggests that some kind of basis dependence is introduced in the construction of~\cite{SL}. Our counter example, in keeping with the formulation of~\cite{SL}, has a similar limitation due to its basis dependence. We remark that any form of basis-dependence of the dynamics raises questions about the linearity of its evolution. We explore this issue in the next section.

\section{Linear maps}\label{linear}

In this section we show how to construct dynamical maps that take a family of states $\{\rho^\s(0)\} \to \{\rho^\s(t)\}$ and lead to Choi matrices that are valid for all states, independent of their basis. We will emphasize the importance of linearity for such maps. We note that in general the transformation $\{\rho^\s(0)\} \to \{\rho^\s(t)\}$ may not define a unique map on the full set of states; the map may be different for different frameworks. The question of linearity and complete-positivity is one that regards one possible map describing the process for the desired family of states.

By definition, a map $\mathcal{B}$ is linear if and only if $\mathcal{B} [\rs_1(0)] + \mathcal{B} [\rs_2(0)] = \mathcal{B} [\rs_1(0) + \rs_2(0)]$ for any two states $\rs_1(0)$ and $\rs_2(0)$. From here on, we will use $\mathcal{B}$ to label linear maps, and keep the label $\Phi$ for maps defined in~\cite{SL}. If one superficially examines the map Eq.~\eqref{mapintheslway} it \emph{seems} linear.

However, note the subtle difference between basis $\{\ket{i}\}$ of Eq.~\eqref{slclass} in comparison to the basis $\{\ket{\chi_i}\}$ of Eq.~\eqref{slclasszd} and $\{\ket{\psi_i}\}$ of Eq.~\eqref{slclass2}. When a map is constructed using a specific basis which depend on the state, the actions of the maps cannot be considered linear. More concretely, consider two different states $\rse_1(0)$ and $\rse_2(0)$, each with vanishing discord, but in two different basis of $\s$. Now construct two maps using the procedure laid out in~\cite{SL}: $\Phi_{1} [\rs_1(0)]$ with the eigenbasis of $\rs_1(0)$, and $\Phi_{2}[\rs_2(0)]$ with a different eigenbasis corresponding to $\rs_2(0)$. These two maps are different from the map constructed from the eigenbasis of the sum $\Phi_{1+2} [\rs_1(0) + \rs_2(0)] \not= \Phi_{1} [\rs_1(0)] + \Phi_{2} [\rs_2(0)]$. From another point of view, each map as constructed in~\cite{SL} would be CP, and their weighted combination should in turn also be CP. But, in general, a state that is a mixture of $\rs_1(0)$ and $\rs_2(0)$ can have non-vanishing discord~\cite{Modi:2011wv}, which would imply that vanishing discord is not a necessary condition for CP maps.

This argument depends on the linearity of maps. One option is to accept the nonlinear definition of the map in Eq.~\eqref{mapintheslway}. However, this choice is very problematic from several points of view. Strictly speaking the CP is referred always to linear maps, and for example Choi's criterion for CP is only valid for linear maps~\cite{choi75}. At the accuracy experimentally available~\cite{Zeilinger, Weinberg, Wineland} quantum theory is well described by a linear maps. Thus, we should always subscribe to linear maps.

Of course, we may wonder about the complete-positivity of the maps given by Eqs.~\eqref{mapproj} and \eqref{CPCE} provided that the operators $E$, $F$ and $M$ are considered as fixed and independent of the state. However, as we have shown, those maps may be CP or not depending on how we fix the basis (e.g. if its elements are chosen to be orthonormal or not). In addition, the price to pay for taking the elements of the operators $E$, $F$ and $M$ to be fixed in Eqs.~\eqref{mapproj} and \eqref{CPCE} is that the maps lose their physical meaning when they are applied for some other states. Concretely, Eq.~\eqref{mapproj} places constraints on $\{\rs_{ij}\}$ due to the requirements for positivity and trace of state in Eq.~\eqref{slclass}. Arbitrary $\s$ states may lead to unphysical $\se$ states, and possibly a meaningless final states for $\s$. A similar thing happens when Eq.~\eqref{CPCE} applied on states that are not invariant under a POVM whose operators are given by the $M$ matrices. If we allow for the operators $E, \; F$ and $M$ to depend on the state which is mapped, then it does not make any sense to apply the Choi criterion, because it is only valid for linear maps.

\subsection{Linear assignment maps}\label{SectionA}

\begin{center}
\begin{figure}[t!]
\includegraphics[scale=0.3]{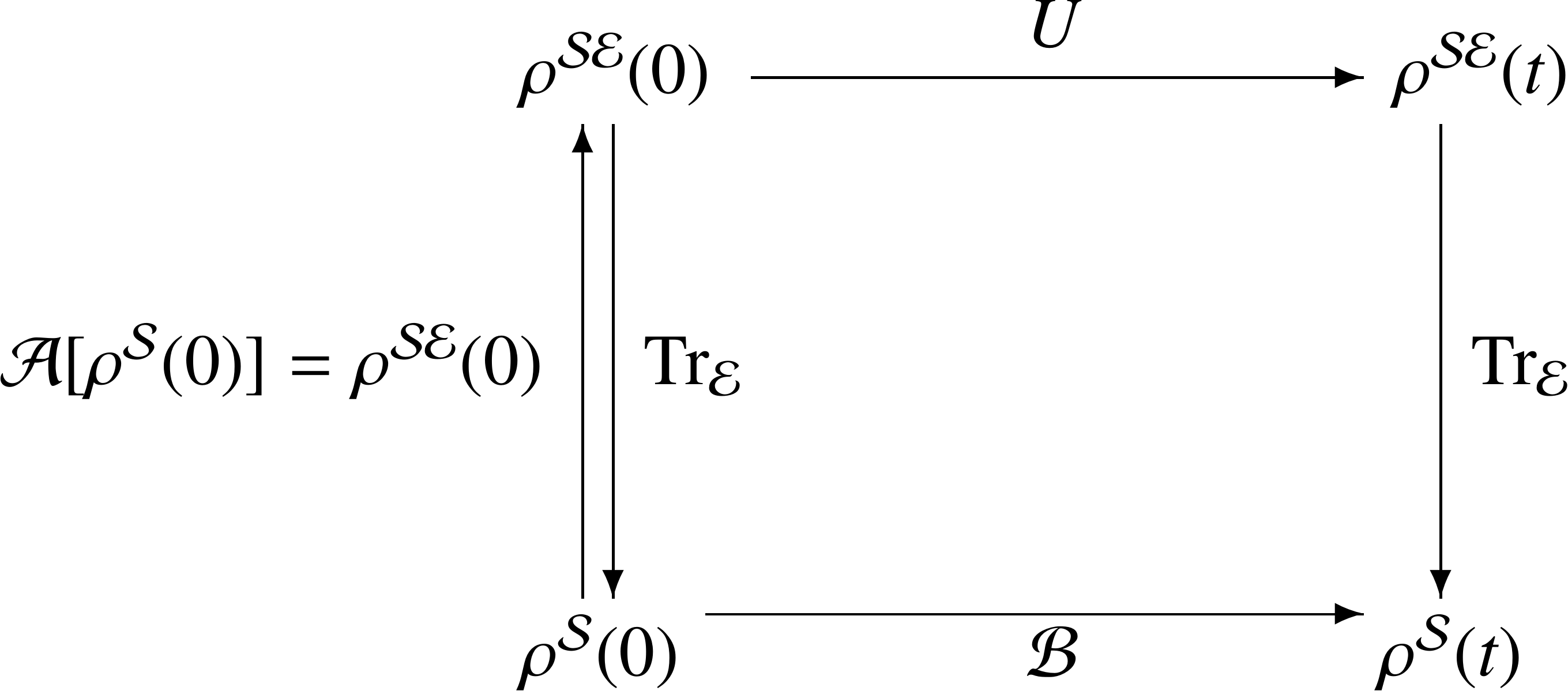}
%
%
\caption{\label{fig1} \emph{Reduced dynamics from total dynamics.} The total state evolves unitarily $\rse(t)=U \rse(0)U^\dag$. The initial state of system $\s$, $\rs(0)=\tre[\rse(0)]$, is mapped to final state $\rs(t)=\mathcal{B}[\rs(0)]$ by the dynamical map $\mathcal{B}$. This process may also be seen as $\rs(0)$ assigned to $\rse(0)$ by the assignment map $\A$ followed by the unitary transformation $U\left( \cdot \right)U^\dagger$, and finally the environment $\e$ is traced by $\tre$.}
\end{figure}
\end{center}

A way to avoid the ambiguities with linearity of dynamical maps is to work with the assignment map formalism~\cite{Pechukas, Alicki, JSS, CTZ, RMA, MSS}. Essentially, to define an assignment map means to make a choice for one of the possible inverses for the partial trace~\footnote{Recall that the partial trace is not a one-to-one map so its inverse is not uniquely defined.}. Thus, the evolution of a reduced state $\rho^\s(0)$ given under an assignment map $\A$ and the unitary evolution $U$ leads to the dynamical map $\mathcal{B}$ of the form:
\begin{gather}\label{Adynamics}
\rho^\s(t)=\mathcal{B}[\rs(0)]=\tre \left\{U \A[\rs(0)] U^\dag \right\}.
\end{gather}
This has been pictorially represented in the diagram in Fig.~\ref{fig1}. The mathematical properties of assignment maps and their corresponding dynamical maps have been thoroughly discussed in~\cite{RMA}. There, it was shown that non-linear assignment maps may lead to violations of well accepted laws, such as the no-cloning theorem.

We stress that the way to construct the map $\Phi$ in Secs.~\ref{SectionSL} and~\ref{SectionCE}, requires an assignment map that depends on the chosen basis of the total state. Note that in Eq.~\eqref{mapproj} the operators $E$ and $F$ depend on the reduced state $\rho^\s = \sum_{i,j} \rho^\s_{ij} \ket{i} \bra{j}$ through the semi-positive definite condition of $\rho^\se = \sum_{i,j} \rho^\s_{ij} \ket{i} \bra{j} \otimes \phi_{ij}^\e$. By a similar argument, in Eq.~\eqref{CPCE} the $E$ operators also depends on the $\rho^\s$.

Using an analogous argument as in~\cite{RMA}, it can be shown that such nonlinear assignments needed to define the states in Eq.~\eqref{slclass} have properties incompatible with the original Choi formulation. To avoid these problems, we will subscribe only to linear assignment maps. Using them, we will show in a different way how vanishing discord is not a necessary condition for CP maps.

Given a linear and positive assignment map $\A$ we ask: \emph{Is the dynamics of $\s$, assigned by $\A$ to $\se$, CP if and only if the assigned state has vanishing discord?} The answer is given by the following two results: The `if' part of the proof is given in the next Theorem, while the `only if' part is shown to be false in the Proposition following the Theorem.

{\bf Theorem.} If the $\se$ state has vanishing discord, there exist a positive linear-assignment map that describes the state of $\se$ for a chosen state of $\s$.

\emph{Proof.} Such an assignment map has the form $\A[\rs] = \sum_i \braket{i \left| \rs \right| i} \ket{i}\bra{i} \otimes \phi^\e_{ii}$. Since $\braket{i \left| \rs \right| i} \ge 0$, one can easily verify that the assignment is positive. A complete proof of the Theorem is given in~\cite{RMA}. \qed

Now, we show that the converse is not true.

{\bf Proposition.} \emph{There are positive linear-assignment maps which mathematically characterize $\se$ states with non-vanishing discord.}

\emph{Proof.} The proof is provided by the state in Eq.~\eqref{ex1}, which has non-vanishing discord. Consider the following CP assignment map
\begin{align}\label{AssingCP}
\A[\rho^\s] = D_{+0}\rho^\s D_{+0}^\dag + D_{+1}\rho^\s D_{+1}^\dag
+\sum_{j=0}^n D_j\rho^\s D_j^\dag,
\end{align}
where the operators of the assignment, $D_i$, are related to the POVM of the counter example, $M_i$, given in Eqs.~\eqref{POVM1}, \eqref{POVM2}, and \eqref{POVM3}, and the operators of $\e$, given in Eqs.~\eqref{OPE1} and \eqref{OPE2}: $D_i = M_i \otimes \sqrt{\phi^\e_i}$. This leads to states with non-vanishing discord. Since the linear assignment map is CP, then the dynamical map $\mathcal{B}$ is also CP.\qed

The two assignment maps above are not consistent, i.e., $\tre\{\A[\rs]\} \ne \rs$ for all $\rs$. The domain of consistency are the states with the form: $\rho^\s = \sum_j p_j \ket{j} \bra{j}$ and $\rho^\s = \frac{p}{3} \left( \ket{0} \bra{0} + \ket{1} \bra{1} + \ket{+} \bra{+} \right) + \sum_{j\geq2} p_j\ket{j} \bra{j}$ respectively for the maps of Theorem and Eq.~\eqref{AssingCP}. This is in perfect agreement with Pechukas' theorem~\cite{Pechukas, JSS}, that a linear-assignment is positive and consistent if and only if there are no initial $\se$ correlations.

We can now rephrase the above results in the following way: \emph{The family of zero discord states $\sum_i \rs_{ii} \ket{i}\bra{i} \otimes \phi^\e_{ii}$, spanned by the fixed operators $\{\ket{i}\bra{i} \otimes \phi^\e_{ii}\}$ and the free parameter $\{\rs_{ii}\}$ is the result of some CP assignment map acting on states in its consistency domain.} However: \emph{one can define a CP assignment map that takes states of $\s$ in its consistency domain to discordant $\se$ states.}

Initial $\se$ states with vanishing discord is a sufficient but not a necessary condition for compatibility with a linear assignment map. Hence in some cases it is possible to describe the dynamics of an initially discordant state using a CP map.

\section{Conclusions}\label{conclusions}
The connection between initial $\se$ correlations and the description of the reduced evolution in terms of CP maps is not as simple as claimed in~\cite{SL}. We show this by using two frameworks; one based on the properties of the global $\se$ state and the other by using assignment maps. We then show that the methods developed in framework (i) and in~\cite{SL} to construct the reduced dynamics from a correlated $\se$ state may lead to nonlinear maps. These maps may take the form of a CP or not CP dynamic regardless of whether the initial $\se$ state has non-vanishing discord. A way to remove this ambiguity is by working with a linear assignment map. In such a case, a positive linear assignment map allows for a $\se$ state with vanishing discord. However, the converse is not true, i.e., there are positive assignment maps that can yield a $\se$ state with non-vanishing discord. Again, we note that such assignment maps are not consistent and have a limited domain of applicability.

Lastly, in an experiment, one does not have access to the environment or any correlations with it. The only accessible  quantities are a preparation device, an unknown process, and a measuring device. Therefore, the only map that ought to be relevant is the one that can be constructed in an experiment. Indeed, the procedure to characterize a dynamical process, known as quantum process tomography, is a function of preparations and measurements. In~\cite{kuah:042113, aharon} it is pointed out that quantum process tomography with initial correlations, quantum or classical, is nontrivial. When dealing with experimental systems, a third framework may be of use~\cite{arXiv:1011.6138}. This framework takes an operational approach, which fixes the initial $\se$ to be a constant, leads to a linear, completely-positive maps, independent of discord in the initial state of $\se$.

\acknowledgements We thank Marco Piani and Daniel Terno for valuable conversations. AB thanks Industry Canada, NSERC  and CIFAR for support. AD is supported by EPSRC (Grant No. EP/H03031X/1), and EU Integrated Project QESSENCE. KM is supported by the John Templeton Foundation, the National Research Foundation, and the Ministry of Education of Singapore. AR thanks financial support from project QUITEMAD S2009-ESP-1594 of the Consejer\'ia de Educaci\'on de la Comunidad de Madrid and MICINN FIS2009-10061. A.B., A.D., and C.R. acknowledge the Benasque Centre for Science where early discussions leading to this publication occurred.

\bibliography{SLCE}

\end{document}